\pdfoutput=1
   \documentclass[fleqn,usenatbib,usedcolumn]{mnras}
   \usepackage[british]{babel}             % British English hyphenation
   \let\la=\lesssim % for less than similar from txfonts, not \la from mnras.cls
   \usepackage[T1]{fontenc}                % Good font encoding
   \usepackage{graphicx}                   % Including figures
\usepackage{siunitx}
\usepackage{mhchem}
\usepackage{pgfplots}
\pgfplotsset{compat=newest}

\DeclareSIUnit\mols{{molec.}\,s^{-1}}

   \volume{\textrm{in press}}
\title[Robotic view of 67P perihelion]{The perihelion activity of comet 67P/Churyumov-Gerasimenko as seen by robotic telescopes}

\author[C. Snodgrass et al.]{%
Colin Snodgrass,$^{1}$\thanks{E-mail: colin.snodgrass@open.ac.uk (CS)}
Cyrielle Opitom,$^{2}$
Miguel de~Val-Borro,$^{3,4}$
Emmanuel Jehin,$^{2}$
\newauthor
Jean Manfroid,$^{2}$
Tim Lister,$^{5}$
Jon Marchant,$^{6}$
Geraint H. Jones,$^{7,8}$
Alan Fitzsimmons,$^{9}$
\newauthor
Iain A. Steele,$^{6}$
Robert J. Smith,$^{6}$
Helen Jermak,$^{6,10}$
Thomas Granzer,$^{11}$
\newauthor
Karen J. Meech,$^{12}$
Philippe Rousselot,$^{13}$
and Anny-Chantal Levasseur-Regourd$^{14}$
\\
\smallskip\\
$^{1}$School of Physical Sciences, The Open University, Walton Hall, Milton Keynes, MK7 6AA, UK\\
$^{2}$Institut d'Astrophysique et de G{\'e}ophysique, Universit{\'e} de Li{\`e}ge, all{\'e}e du 6 Ao{\^u}t 17, B-4000 Li{\`e}ge, Belgium\\
$^{3}$Department of Astrophysical Sciences, Princeton University,
Princeton, NJ 08544, USA\\
$^{4}$NASA Goddard Space Flight Center, Astrochemistry Laboratory, Code 691.0,
Greenbelt, MD 20771, USA\\
$^{5}$Las Cumbres Observatory Global Telescope Network, 6740 Cortona Drive Ste. 102, Goleta, CA 93117, USA\\
$^{6}$Astrophysics Research Institute, Liverpool John Moores University, Liverpool, L3 5RF, UK\\
$^{7}$Mullard Space Science Laboratory, University College London, Holmbury St. Mary, Dorking, Surrey RH5 6NT, UK\\
$^{8}$The Centre for Planetary Sciences at UCL/Birkbeck, Gower Street, London WC1E 6BT, UK\\
$^{9}$Astrophysics Research Centre, School of Mathematics and Physics, Queen's University Belfast, BT7 1NN, UK\\
$^{10}$Physics Department, Lancaster University, Lancaster, LA1 4YB, UK\\
$^{11}$Leibniz-Institut f{\"u}r Astrophysik Potsdam,
An der Sternwarte 16, 14482 Potsdam, Germany\\
$^{12}$Institute for Astronomy, 2680 Woodlawn Drive, Honolulu, HI 96822, USA\\
$^{13}$University of Franche-Comt\'e, Observatoire des Sciences de l'Univers THETA, Institut UTINAM - UMR CNRS 6213, BP 1615, \\
25010 Besan\c con Cedex,France\\
$^{14}$LATMOS-IPSL; UPMC (Sorbonne Univ.), BC 102, 4 place Jussieu, 75005 Paris, France
}

\date{Accepted XXX. Received YYY; in original form ZZZ}

\pubyear{2016}

\begin{document}
\label{firstpage}
\pagerange{\pageref{firstpage}--\pageref{lastpage}}
\maketitle

\begin{abstract}
Around the time of its perihelion passage the observability of 67P/Churyumov-Gerasimenko from Earth was limited to very short windows each morning from any given site, due to the low solar elongation of the comet. The peak in the comet's activity was therefore difficult to observe with conventionally scheduled telescopes, but was possible where service/queue scheduled mode was possible, and with robotic telescopes. We describe the robotic observations that allowed us to measure the total activity of the comet around perihelion, via photometry (dust) and spectroscopy (gas), and compare these results with the measurements at this time by Rosetta's instruments. The peak of activity occurred approximately two weeks after perihelion. The total brightness (dust) largely followed the predictions from \citet{Snodgrass2013}, with no significant change in total activity levels from previous apparitions. {The CN gas production rate matched previous orbits near perihelion, but appeared to be relatively low later in the year}.
\end{abstract}


%%%%%%%%%%%%%%%%%%%%%%%%%%%%%%%%%%%%%%%%%%%%%%%%%%

%%%%%%%%%%%%%%%%% BODY OF PAPER %%%%%%%%%%%%%%%%%%

\section{Introduction}

A large world-wide campaign of ground-based observations supported the { European Space Agency's unique} Rosetta mission, { the first spacecraft to orbit a comet, which followed 67P/Churyumov-Gerasimenko (hereafter 67P) from 2014 to 2016 as it passed through perihelion}. 
The campaign had the dual purpose of providing large scale context for Rosetta, by measuring total production rates and observing the coma and tails beyond the spacecraft's orbit, and allowing comparison between 67P and other comets. 
Predictions for the total dust activity of the comet were made by \citet{Snodgrass2013}, based on observations from previous orbits, and observations in the pre-landing phase of the mission showed the comet to be following these predictions \citep{Snodgrass2016}.
This implies that there is little change from orbit-to-orbit in 67P, and that results from Rosetta can be more generally applied. 
A simple thermophysical model (balancing the sublimation needed to produce the observed dust coma with the input solar irradiance -- e.g. \citet{Meech}) was able to describe most observations presented by \citet{Snodgrass2013}, but { underestimated the peak brightness relative to the data in} the region around the perihelion passage. The peak in activity was also an important opportunity to measure the total gas production of the comet, after deep searches with large aperture telescopes could only produce upper limits to emissions in the 2014 observing window \citep{Snodgrass2016}. Good coverage of the perihelion passage was therefore a priority for the ground-based observation campaign, despite the challenging observing geometry.

67P was in Southern skies during the 2014 observing window (February -- November), and slowly brightened as it approached the Sun from $\sim4 - 3$ AU, but still required large aperture telescopes to observe. After a gap in coverage enforced by low solar elongation between December 2014 and April 2015 the comet was briefly observable from the Southern hemisphere before reaching declination of +24\degr{} around the time of its perihelion passage (August 2015). Throughout the perihelion period the phase angle was around 30 degrees, and the solar elongation varied from 30 to 90 degrees, with the comet observable only in morning twilight for the majority of the time. At this time smaller aperture robotic telescopes were better able to follow the comet than the larger facilities used in 2014. 

In this paper we describe observations taken as part of an International Time Programme (ITP) using Canary Island telescopes, the 2m Liverpool Telescope on La Palma and the 1.2m STELLA imaging telescope on Tenerife, and with the specialist comet observing 0.6m telescope TRAPPIST, at La Silla in Chile, and the robotic telescopes operated by Las Cumbres Observatory Global Telescope Network (LCOGT). Observations in the months around perihelion (13 August 2015) are included in the current work, corresponding to the period when the peak in activity was observed, while the comet was at heliocentric distances $1.2 < r < 2$ AU.
% r < 2 AU covers April - December 2015.
The following section describes the observations from each telescope, while
section~\ref{sec:activity} reports the total activity measurements derived. We
discuss the implications of these results, and compare them with earlier
observations and Rosetta measurements, in section~\ref{sec:discussion}.

\section{Observations}

\begin{table}
  \centering
  \caption{Log of observations described in this paper, together with
  heliocentric and geocentric distances ($r$ \& $\Delta$, AU) and solar phase
angle ($\alpha$, degrees). Dates are all 2015, format MM-DD.dd. TRAPPIST data. {Full table available online, first 5 rows given as an example.}}
  \label{tab:obs1}
  \begin{tabular}{lllccc} 
    \hline
    UT date & Tel./Inst. & $N \times$ $t_{\rm exp}$ filter  & $r$ & $\Delta$ & $\alpha$\\
    \hline
    04-18.41 & TRAPPIST &  2 $\times$ 240s Rc & 1.83 & 2.64 & 15.6 \\
    04-25.42 & TRAPPIST &  3 $\times$ 180s Rc & 1.78 & 2.56 & 17.2 \\
    04-29.42 & TRAPPIST &  2 $\times$ 180s Rc & 1.75 & 2.51 & 18.1 \\
    05-04.42 & TRAPPIST &  1 $\times$ 180s Rc & 1.71 & 2.45 & 19.2 \\
    05-05.42 & TRAPPIST &  3 $\times$ 180s Rc & 1.71 & 2.44 & 19.5 \\
... & ... & ... & ... & ... & ... \\
     \hline
  \end{tabular}
\end{table}

We summarise the observations around perihelion in Table~\ref{tab:obs1}, and describe them in more detail in the following sub-sections.

\subsection{TRAPPIST}
\label{sec:trappist} % used for referring to this section from elsewhere

TRAPPIST (TRAnsiting Planets and Planetesimals Small Telescope) is a 60-cm robotic telescope installed in 2010 at La Silla observatory \citep{TRAPPIST}. The telescope is equipped with a $\mathrm{2K\times2K}$ thermoelectrically cooled FLI Proline CCD camera with a field of view of 22\arcmin x22\arcmin. We binned the pixels 2 by 2 and obtained a resulting plate scale of 1.3\arcsec /pixel. The telescope is equipped with a set of narrow-band filters designed for the observing campaign of comet Hale-Bopp \citep{Farnham2000} isolating the emission of OH, NH, CN, $\mathrm{C_3}$, $\mathrm{C_2}$ and emission free continuum regions at four wavelengths. A set of broad band B, V, Rc, and Ic Johnson-Cousin filters is also mounted on the telescope. We observed the comet once or twice a week from April 18, 2015 to the end of the year, with broad band filters. Exposure times ranged from 120 to 240 s. Technical problems and bad weather prevented observations for some days
around perihelion. Between August 22, 2015 and September 12, 2015 we were able to detect the CN emission using narrow band filters. The $\mathrm{C_2}$ was also detected but the SNR was not sufficient to derive reliable gas production rates. We could not detect the OH, NH, or $\mathrm{C_3}$ emission. { Upper limits on the production rates for gas species other than CN have not yet been derived, but will be presented as part of a more in-depth study of using TRAPPIST and VLT data (Opitom et al., in prep.) }.

Calibration followed standard procedures using frequently updated master bias, flat and dark frames. The removal of the sky contamination and the flux calibration were performed as described in \citet{Opitom2015}. Median radial profiles were extracted from each image and dust contamination was removed from the CN profiles. Observations in the Rc broad band filter were used to derive total R-band magnitudes at 10,000 km. We also derived the $Af\rho$ at 10,000 km and corrected it from the phase angle effect using a function which is a composite of two different phase functions from \citet{Schleicher1998} and \citet{Marcus2007}. From the observations in the CN narrow band filter, we derived CN production rates. The CN fluxes were converted into column density and we adjusted a Haser model \citep{Haser1957} on the profiles to derive the production rates. The model adjustment was performed around a physical distance of 10,000 km from the nucleus to avoid PSF and seeing effects around the optocentre and low signal-to-noise ratio at larger nucleocentric distances. We used a constant outflow velocity of 1 km/s as assumed by \citet{Ahearn1995}, together with their scale lengths scaled as $r^2$, $r$ being the heliocentric distance.

\subsection{Liverpool Telescope}
\label{sec:LT} % used for referring to this section from elsewhere

The 2m Liverpool Telescope (LT) was one of the first fully robotic professional
telescopes, and has been in operation at Roque de los Muchachos Observatory on La
Palma since 2003. It was built and is operated by Liverpool John Moores
University, and is equipped with an array of instruments (imagers,
spectrographs and polarimeters) that can be quickly switched between during the
night \citep{Steele2004}. As one of the larger telescopes that could regularly
observe 67P around perihelion, we proposed to use it primarily for
spectroscopy, to study the dust colours (continuum slope) and gas emission
bands. As the existing long-slit spectrograph, SPectrograph for the Rapid
Acquisition of Transients \citep[SPRAT;][]{Piascik2014},
covers red wavelengths where cometary gasses have only weaker emissions, the LT
team proposed the creation of a new low resolution ($R\sim330$) 
blue/UV { (320--630 nm)} sensitive spectrograph for the 67P
monitoring programme. The rapid design, construction and commissioning of this
instrument, the LOw-cosT Ultraviolet Spectrograph (LOTUS), enabled us to
observe the stronger CN band at 388 nm \citep{lotus}.

Observations of 67P using LOTUS began on 2015-09-05 and continued until the
comet had faded too far for UV spectroscopy with the LT\@. LOTUS was designed
with a two-width slit, the longer and narrower part slit of 2.5''$\times$95'' being optimised for comet
observations, while the wider 5''$\times$25'' slit allowed observations of spectrophotometric
standard stars to measure the instrument response.
The CCD pixels were binned \num{4 x 4} to obtain a spatial pixel
scale of 0.6\arcsec/pixel.

LOTUS spectroscopic data were reduced with the routine pipeline to produce
science frames. This pipeline is based on the FRODOSpec reduction pipeline
\citep{2012AN....333..101B} and is similar to that of other long slit
instruments.  First the bias and dark frames were subtracted and the wavelength
calibration carried out.  The pipeline automatically aligns the dispersion
direction in the two-dimensional frames with rows of the array to produce
wavelength calibrated spectra.  Three 300-s comet spectra obtained for each
epoch were median-combined and extracted by summing the flux over an aperture
along the slit.  Several frames with clean sky background were observed at the
same airmass to perform the sky subtraction for each combined frame.  Finally,
the spectra of the comet were corrected for atmospheric extinction and flux
calibrated with observations of standard star using standard IRAF techniques.
The continuum in the comet spectra caused by sunlight reflection off
the dust was removed using the spectrum of the solar analogue HD29641 that was
observed in the beginning of the observing period.

From the LOTUS observations we derived CN production rates
using a Haser spherically symmetric model \citep{Haser1957} that is also used
for the analysis of the TRAPPIST photometric data described
in~\ref{sec:trappist}.  We included the photo-production and dissociation of
molecules in the coma with parent and daughter scale-lengths and fluorescence
efficiencies taken from \citet{2010AJ....140..973S}.  With this model the
column density in a circular area of the observed species is proportional to the
measured flux of the emission band, from which the production rate
can be derived directly.

There were also images of the comet collected with the LT, using its IO:O
camera \citep{steele2014}. These include the acquisition images taken to
robotically acquire the comet onto the slit of LOTUS, and some additional
deeper images. Images were mostly taken in the SDSS-$r$ filter, with additional
sets in $griz$ taken to measure the colour of the coma in the weeks closest to
perihelion. All images taken with the LT were processed using the IO:O
pipeline, which performs bias subtraction, flat fielding, and photometric
calibration.

\subsection{STELLA}
\label{sec:stella} % used for referring to this section from elsewhere

The STELLA telescopes are a pair of 1.2m telescopes at the Teide observatory on the island of Tenerife, built and operated by the  Leibniz-Institut f{\"u}r Astrophysik Potsdam (AIP) in collaboration with the Instituto de Astrof{\`i}sica de Canarias (IAC). The pair of telescopes have complementary instrumentation -- a wide-field imager on one telescope and high resolution spectrograph on the other -- and were built with monitoring of stellar activity in cool stars in mind. Further details on the telescopes can be found in the papers by \citet{STELLA1,STELLA2}.

We used the imaging telescope to perform imaging in an SDSS-$r$ filter on every possible night, and attempted $griz$ filter observations every 10 nights. The instrument, the Wide Field STELLA Imaging Photometer (WiFSIP), has a 22' field of view and 0.32"/pix pixel scale, using a single 4k CCD. The STELLA telescope TCS does not allow tracking of moving (solar system) objects, and expects a fixed RA and dec for each target. Observing blocks (OBs) are created and submitted to the queue using a java tool. In order to interact with this system, creation of the OBs was scripted to produce one block per night with appropriate start and end time constraints, the correct position, and a short enough exposure time that the comet would not move more than 0.5" during the exposure (and therefore stay within the seeing disc). The number of exposures was scaled to have an approximately fixed OB length (10 minutes in the near-perihelion period). These OBs were then inserted directly into the telescope queue.

Data were taken robotically and automatically reduced using the STELLA pipeline, which also determines individual frame zeropoints by matching field stars with the PPMXL catalogue \citep{2010AJ....139.2440R}. This is based on USNO-B1.0 and 2MASS catalogues; transformations from \citet{2008MNRAS.384.1178B} and \citet{2005AJ....130..873J} are used to give zeropoints in the SDSS-like filters. The resulting absolute calibration is internally consistent, as can be seen by the smooth night-to-night variation, and gives a good match in the $r$-band to other total brightness values from other telescopes, but the filter-to-filter zeropoints are not well calibrated, and therefore further calibration is required to measure the colour of the comet using the STELLA data.  

The resulting frames for each night were shifted and stacked, based on the predicted motion of the comet, to produce one median image per filter and night. These were inspected visually to confirm that the comet was detected, and remove any nights where the comet fell on top of a star or there were issues with the data quality. There were occasionally problems with the telescope focus, which is  automatic but did not always perform well at the low elevation the telescope needed to point to for observations of 67P. Badly defocussed images were rejected. The total comet brightness was then measured within various apertures -- here we report the brightness within $\rho$ = 10,000 km at the distance of the comet.

\subsection{LCOGT}
\label{sec:lcogt} % used for referring to this section from elsewhere

Las Cumbres Observatory Global Telescope Network (LCOGT) is a global network of robotic telescopes designed for the study of time-domain phenomena on a variety of timescales. The LCOGT Network incorporates the two 2m Faulkes Telescopes (very similar to the LT described in Section~\ref{sec:LT}) and nine 1m telescopes deployed at a total of five locations around the world, and have been operating as a combined network since May 2014. The LCOGT Network is described in more detail in \citet{Brown2013} and the operation of the network is described in \citet{Boroson2014}.

Due to the visibility of 67P, we started observations on 2015-08-07 with the 2m Faulkes Telescope North (FTN) on Haleakala, Maui, Hawaii using the \textit{fs02} instrument. This instrument is a Spectral Instruments 600 camera using a Fairchild $4096\times4096$ pixel CCD486 CCD which was operated in bin $2\times2$ mode to give a pixel scale of 0.3"/pixel and a field of view of $10\arcmin \times 10\arcmin$. Observations were primarily conducted in SDSS-$g'$ and SDSS-$r'$ but a series of $g'r'i'z'$ observations were taken on 7 nights between 2015-09-03 and 2015-09-21.

Observations with the 1m network started on 2015-12-08 and continued through until 2016-03-26 -- in this paper we describe data taken up to the end of December 2015. Observations were obtained from the LCOGT sites at McDonald Observatory (Texas; 1 telescope), Cerro Tololo (Chile; 3 telescopes), Sutherland (South Africa; 2 telescopes) and Siding Spring Observatory (Australia; 1 telescope) and were all obtained in SDSS-$r'$. Two different instrument types were used; the first one uses a SBIG STX-16803 camera with a Kodak KAF-16803 CCD with $4096\times4096$ $9\micron$ pixels which was operated in bin $2\times2$ mode to give a pixel scale of 0.464"/pixel and a field of view of $15.8\arcmin \times 15.8\arcmin$. The other instrument type was the LCOGT-manufactured \textit{Sinistro} camera using a Fairchild $4096\times4096$ pixel CCD486 CCD which was operated in bin $1\times1$ mode to give a pixel scale of 0.387"/pixel and a field of view of $26.4\arcmin \times 26.4\arcmin$. 

All data were reduced with the LCOGT Pipeline based on \textsc{ORAC-DR} (\cite{Jenness2015} and also described in more detail in \cite{Brown2013}) to perform the bad-pixel masking, bias and dark subtraction, flat-fielding, astrometric solution and source catalog extraction. In order to produce a more consistent result and to allow the use of photometric apertures that are a fixed size at the distance of the comet (and therefore of variable size on the CCD), we elected to resolve the astrometric and photometric (zeropoint determination) solution for all the data using a custom pipeline that operated on the results of the LCOGT Pipeline. 

This comet-specific pipeline makes use of \textsc{SExtractor} \citep{SExtractor} and \textsc{scamp} \citep{SCAMP} to produce a source catalog and solve for the astrometric transformation from pixel co-ordinates to RA, Dec. The UCAC4 catalog \citep{Zacharias2013UCAC4} was used by \textsc{scamp} to determine the astrometric solution and also by the pipeline to determine the zeropoint between the instrumental magnitudes and the magnitude for cross-matched sources in the UCAC4 catalog. Outlier rejection was used to eliminate those cross-matches with errant magnitudes (in either the CCD frame or the UCAC4 catalog) and this process was repeated until no more cross-matches were rejected. In a small number of cases, readout/shutter problems with the SBIG cameras caused part of the image to receive no light and the zeropoint determination failed in these cases. The frames were excluded from the analysis.

The comet magnitudes were then measured through photometric aperture centered on the predicted position and with a radius corresponding to 10,000\,km at the time of observation, taking into account the appropriate pixel scale of the instrument used and the Earth--comet distance. The predicted position and distance were interpolated for the midpoint of the observation in ephemeris output produced by the JPL \textsc{horizons} system \citep{horizons}.
%Figures are referred to as e.g. Fig.~\ref{fig:example_figure}, and tables as
%e.g. Table~\ref{tab:example_table}.

\section{Activity measurements}
\label{sec:activity}

\subsection{Total brightness}

\begin{figure}
	\includegraphics[width=\columnwidth,trim=40 35 85 80,clip]{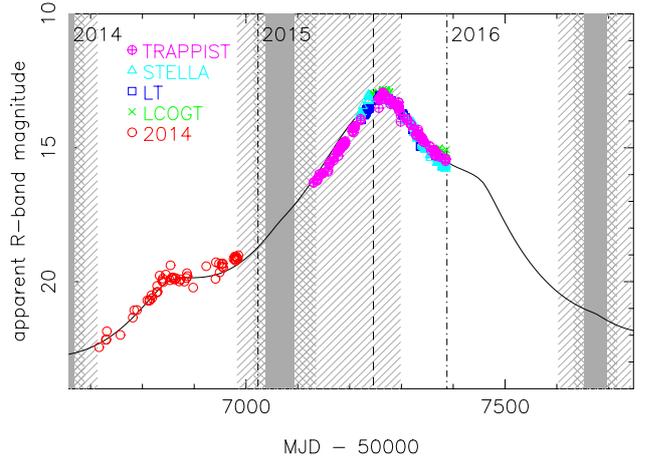}
    \caption{Prediction vs measurements of total $R$-band magnitude within $\rho$ = 10,000 km. Hatched, cross hatched and solid shading shows periods when the solar elongation is less than 50, 30 and 15 degrees respectively. The vertical dashed line marks perihelion.}
    \label{fig:predict-full}
\end{figure}

\begin{figure}
	\includegraphics[width=\columnwidth,trim=40 35 85 80,clip]{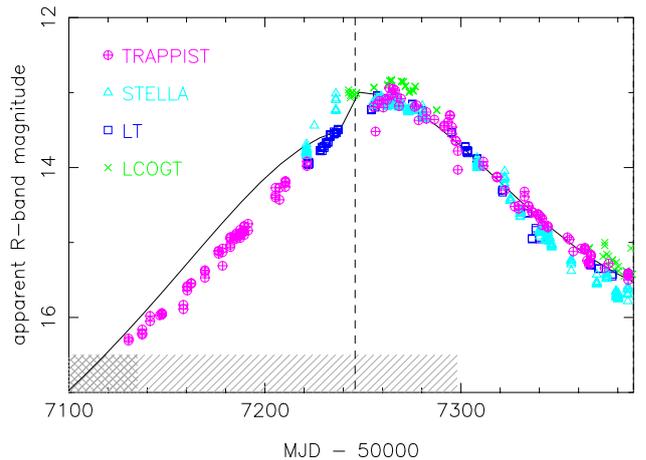}
    \caption{Same as Fig.~\ref{fig:predict-full}, showing zoom in on 2015 period around perihelion. Here the Solar elongation is shown by shading at bottom of the plot only, for ease of seeing data points.}
    \label{fig:predict-zoom}
\end{figure}

Figure \ref{fig:predict-full} recreates the predicted apparent brightness of the comet from \citet{Snodgrass2013} (their Figure 10), with photometry from 2014 \citep{Snodgrass2016} and this work overlaid. It can be seen that the total brightness of the comet, as measured in the $R$-band within a $\rho$ = 10,000 km radius aperture, is in very good agreement with predictions throughout the current apparition. In this figure, and all subsequent ones, we have converted SDSS $r'$ filter photometry to $R$-band for ease of comparison with previous VLT FORS photometry and the \citet{Snodgrass2013} predictions. We use the conversion from Lupton\footnote{http://cas.sdss.org/dr6/en/help/docs/algorithm.asp?key=sdss2UBVRIT}, $R = r - 0.1837(g-r) - 0.0971$, together with the $(g-r) = 0.62 \pm 0.04$ colour of the comet measured with the LT (see section \ref{sec:colours} below). We show a zoom in on the 2015 data ($r < 2$ AU) in Fig.~\ref{fig:predict-zoom}, which shows a number of features. Firstly, the good agreement between the photometry with different telescopes and filters following the conversion to $R$-band is clear. Secondly, there is an obvious offset in the peak brightness post perihelion, and a strong asymmetry -- the apparent magnitude of the comet is brighter at the same distances post-perihelion than pre-perihelion. Finally, it is also apparent that the photometry pre-perihelion is consistently fainter than the predicted curve, although it follows the same trend. Following the method used in \citet{Snodgrass2016}, we find that this implies a drop in total activity in this period of $37\pm9$\% relative to previous orbits, but we caution that the empirical prediction is only meant to be approximate. The step in the prediction curve between pre- and post-perihelion models, for example, is not a real feature. The very good match to the prediction post-perihelion suggests that there is no significant difference in total activity levels this apparition (difference in flux is $3\pm9$\%), and the smooth curve through the post-perihelion peak implies that the mismatch pre-perihelion is probably due to the simplification in the models (which are simple power law fits to heliocentric distance pre- and post-perihelion).

\begin{figure}
	\includegraphics[width=\columnwidth,trim=40 35 85 80,clip]{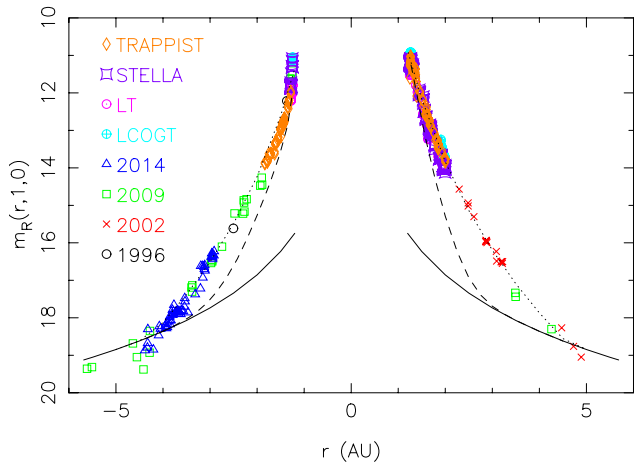}
    \caption{Total $R$-band magnitude within $\rho$ = 10,000 km, corrected to unit geocentric distance and zero degrees phase angle, against heliocentric distance. The solid line shows the predicted magnitude of the inactive nucleus, the dashed line a prediction based on the expected total water prediction rate, and the dotted line an empirical prediction based on photometry from previous orbits \citep[see][for details]{Snodgrass2013}}
    \label{fig:HLC}
\end{figure}

\begin{figure}
	\includegraphics[width=\columnwidth,trim=40 35 85 80,clip]{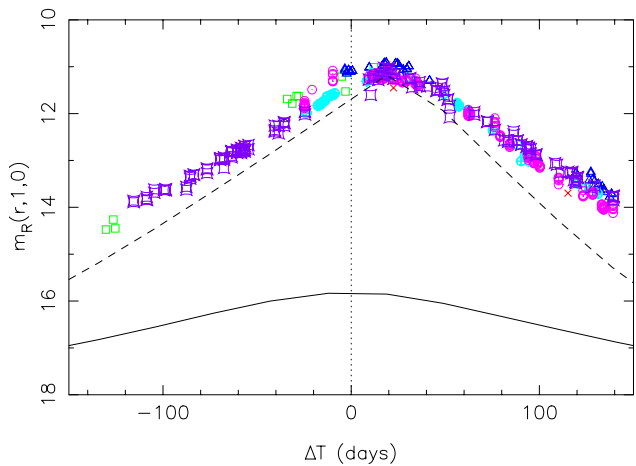}
    \caption{Same as Fig.~\ref{fig:HLC}, but plotted against time from perihelion (days) and showing only the near-perihelion period ($\pm$ 150 days).}
    \label{fig:timeplot}
\end{figure}

In figs.~\ref{fig:HLC} and \ref{fig:timeplot} we plot the same photometry reduced to unit geocentric distance and zero phase angle, as a function of heliocentric distance and time from perihelion respectively. We assume a linear phase function with $\beta = 0.02$ mag. deg$^{-1}$., which is a good approximation for cometary dust \citep[see discussion in][]{Snodgrass2016}. These plots show the data and models from \citet{Snodgrass2013,Snodgrass2016} as well as the perihelion data, and show the consistency between the brightness this apparition and previous orbits.

The total dust activity can also be expressed using the commonly used $Af\rho$
parameter \citep{Ahearn84}. We find that the comet peaked with $Af\rho \sim 400$ cm, or $Af\rho \sim 1000$ cm including a correction to zero phase angle, in the weeks after perihelion. Values for $Af\rho$ are given alongside
the measured magnitudes in table~\ref{tab:phot}.
%The $Af\rho$ curve measured by TRAPPIST is shown in Fig.~\ref{fig:trappist-afrho}.

\subsection{Colour of the coma}
\label{sec:colours}

Using the observations from the LT, we derive average colours for the coma in SDSS bands:
$(g-r) = 0.62 \pm 0.04$, $(r-i) = 0.11 \pm 0.03$, and $(i-z) = -0.45 \pm 0.04$. These are redder, approximately the same, and bluer than the Sun 
{ [$(g-r)_\odot=0.45$, $(r-i)_\odot=0.12$, $(i-z)_\odot=0.04$ -- \citet{Holmberg2006}]}, 
respectively. This follows the general pattern seen in earlier data, and observations from Rosetta, of the spectral slope being bluer at longer wavelengths \citep{Snodgrass2016,Capaccioni2015Sci,Fornasier-AA}, but the extremely blue $(i-z)$ colour is surprising. The calibration of this photometry is based on pipeline results, and the colour term found for calibration in the $i$ and $z$ filters is relatively uncertain, which could contribute some uncertainty -- the individual measurements in $(r-i)$ and $(i-z$) are more variable than the $(g-r)$ colour, which is very stable around the average value for all epochs. Consequently, we have more confidence in the $(g-r$) colour, but regard the other colours as requiring confirmation based on direct calibration of the frames.  This will be done with the forthcoming public release of all sky photometric catalogues in these bands (e.g. Pan-STARRS 1), which will allow direct calibration against field stars in each frame, as part of a more detailed study on the long term evolution of the colours of the coma, { including $griz$ photometry from the LT, STELLA and LCOGT telescopes} (to be published in a future paper). Finally, we  note that the $g$-band brightness contains both dust continuum and C$_2$(0-0) band emission so that the intrinsic dust $(g-r)$ colour will be slightly redder than measured, but our LOTUS spectra show this to be extremely weak.

\subsection{Gas production}

 \begin{figure}
   \centering
   \includegraphics[width=\columnwidth]{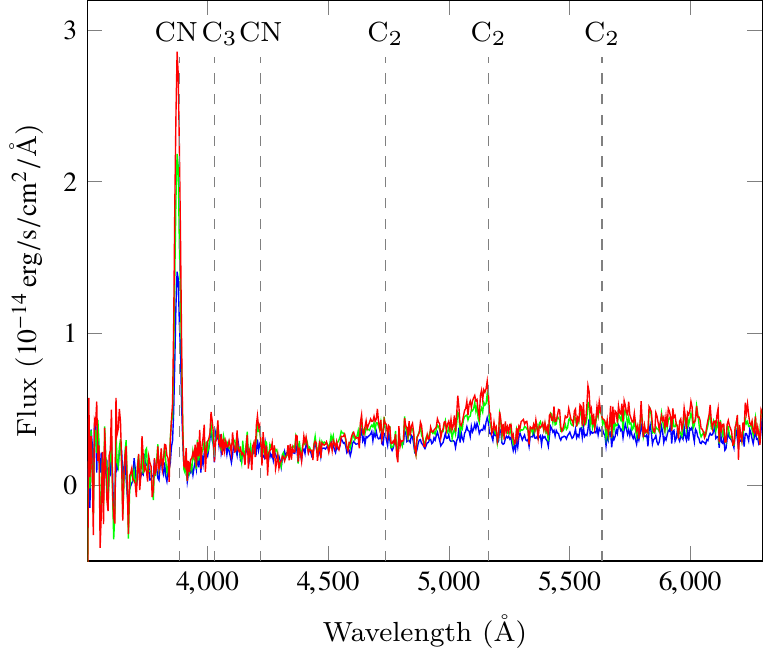}
%   \caption{Average spectrum of observations obtained on 2015-09-05 with
%      LOTUS divided by the spectrum of the solar analogue HD28641.  The total
%      exposure time is 900 s and the sky background has been subtracted.
% }
\caption{Sky-subtracted spectra of comet 67P obtained on September 5 with LOTUS. The  spectra are extracted at three different apertures centred on the comet nucleus with diameters 8.4" (blue line), 16.8" (green line) and 33.6" (red line).}
 \label{fig:lotus_spectrum}
 \end{figure}

We can also assess the total activity of the comet in terms of gas production. From the ground it is difficult to assess the production rates of H$_2$O, CO or CO$_2$ that dominate the coma, so we have to make the assumption that more easily detected species are representative of the total gas production. Rosetta results show that the picture is more complicated \citep[e.g.][]{LeRoy2015,LuspayKuti2015}, but a first order assumption that more CN, for example, implies more total gas is probably still valid. 

 We show an example LOTUS spectrum in Figure~\ref{fig:lotus_spectrum}.
Emission from CN was detected with LOTUS at \SI{3880}{\angstrom}, as well as several spectral
features due to \ce{C2} emission at \SIlist{4738;5165;5635}{\angstrom}. The strongest feature, CN, was detectable from perihelion until the end of 2015, at nearly 2 AU.
We were also able to detect CN at a handful of epochs close to perihelion using narrowband photometry with TRAPPIST, but the comet was too faint (and too low elevation when viewed from Chile) to follow its production rate over an extended period via photometry. 

\begin{figure}
  \centering
  \begin{tikzpicture}
    \begin{semilogyaxis}[
          only marks,
          xlabel={$r_\mathrm{h}$ [AU]},
          ylabel={$Q_\mathrm{CN}$ [\si{\mols}]},
          ymin=1e23
    ]
  \addplot[red,mark=*,error bars/.cd,y dir=both,y explicit]
        coordinates {
          (1.275, 8.736e+24) +- (0., 6.98e+23)
          (1.290, 8.930e+24) +- (0., 7.39e+23)
          (1.308, 7.616e+24) +- (0., 7.79e+23)
          (1.374, 5.474e+24) +- (0., 8.19e+23)
          (1.411, 2.829e+24) +- (0., 8.60e+23)
          (1.446, 3.231e+24) +- (0., 9.00e+23)
          (1.527, 2.058e+24) +- (0., 9.40e+23)
          (1.581, 1.136e+24) +- (0., 9.81e+23)
          (1.616, 2.371e+24) +- (0., 1.02e+24)
          (1.630, 9.444e+23) +- (0., 9.06e+23)
          (1.644, 5.896e+23) +- (0., 5.10e+23)
          (1.659, 9.822e+23) +- (0., 9.14e+23)
          (1.799, 3.421e+23) +- (0., 3.18e+23)
          (1.853, 1.396e+24) +- (0., 1.22e+24)
        };
  \addplot[blue,mark=*,error bars/.cd,y dir=both,y explicit]
        coordinates {
          (1.25, 6.72e+24) +- (0., 6.4e+23)
          (1.25, 7.77e+24) +- (0., 8.18e+23)
          (1.26, 1.00e+25) +- (0., 9.97e+23)
          (1.29, 8.45e+24) +- (0., 9.25e+23)
          (1.30, 7.49e+24) +- (0., 9.01e+23)
        };
  \addplot[green,mark=o]
        coordinates {
          (1.355, 8.32e+24)
          (1.355, 8.71E+24)
		  (1.358, 4.47E+24)
		  (1.358, 4.07E+24)
		  (1.361, 1.07E+25)
		  (1.361, 1.05E+25)
		  (1.514, 6.61E+24)
		  (1.675, 5.13E+24)
		  (1.854, 2.69E+24)
		  (1.862, 2.75E+24)
		  (1.300, 6.31E+24)
		  (1.300, 6.17E+24)
		  (1.374, 1.12E+25)
		  (1.377, 6.31E+24)
		  (1.648, 3.72E+24)
        };
    \end{semilogyaxis}
  \end{tikzpicture}
  \caption{Post-perihelion CN production rates in comet 67P as a function of
    heliocentric distance with 1-$\sigma$ uncertainties.  Red symbols are
    measurements by LOTUS, blue from TRAPPIST, and green are points from previous orbits from \citet{Schleicher2006}.}
\label{fig:cn_production}
\end{figure}
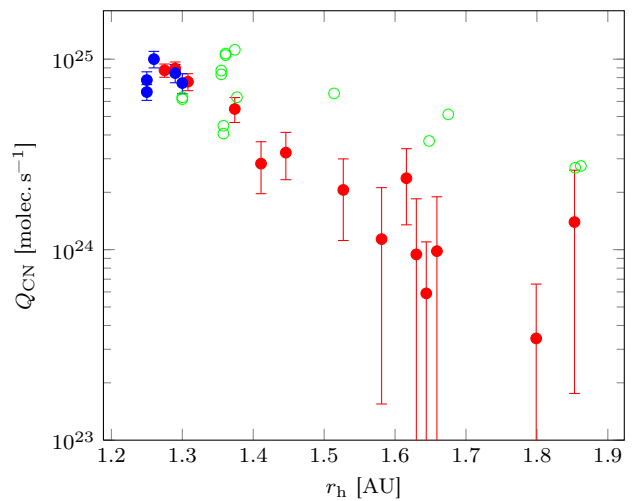

Figure~\ref{fig:cn_production} shows the \ce{CN} production rates as a function
of heliocentric distance obtained with LOTUS and TRAPPIST. The measurements from the different telescopes and techniques are in good agreement during the period when they overlap. The total CN-production rate falls off slowly with increasing heliocentric distance, noticeably different from the steep decrease in total brightness from the dust photometry described above.

\begin{table}
% info from TRAPPIST and LOTUS, sensible columns TBC.
% uncertainties need to be adjusted
	\centering
	\caption{CN production rates measured by TRAPPIST and LOTUS.}
	\label{tab:CN}
	\begin{tabular}{llcc
                        S[
                          separate-uncertainty=true,
                          table-format=1.2(2)e2,
                        ]}
		\hline
UT Date      & Tel./inst. & $r$  & $\Delta$ & {Q(CN)}           \\
   2015          &            & (AU) & (AU)     & {\si{\mols}}      \\
		\hline
08-22.43 & TRAPPIST   & 1.25 & 1.77     & 6.72 +- 0.64E24  \\
08-24.42 & TRAPPIST   & 1.25 & 1.77     & 7.77 +- 0.82E24  \\
08-29.42 & TRAPPIST   & 1.26 & 1.77     & 1.00 +- 0.10E25 \\
09-11.41 & TRAPPIST   & 1.29 & 1.78     & 8.45 +- 0.93E24  \\
09-12.41 & TRAPPIST   & 1.30 & 1.78     & 7.49 +- 0.91E24  \\
09-05.24 & LT/LOTUS   & 1.28 & 1.77     & 8.74 +- 0.70e24  \\
09-10.24 & LT/LOTUS   & 1.29 & 1.78     & 8.93 +- 0.74e24  \\
09-15.23 & LT/LOTUS   & 1.31 & 1.78     & 7.62 +- 0.78e24  \\
09-30.24 & LT/LOTUS   & 1.37 & 1.80     & 5.47 +- 0.82e24  \\
10-07.23 & LT/LOTUS   & 1.41 & 1.80     & 2.83 +- 0.86e24  \\
10-13.24 & LT/LOTUS   & 1.45 & 1.81     & 3.23 +- 0.90e24  \\
10-26.22 & LT/LOTUS   & 1.53 & 1.81     & 2.06 +- 0.94e24  \\
11-03.25 & LT/LOTUS   & 1.58 & 1.80     & 1.14 +- 0.98e24  \\
11-08.23 & LT/LOTUS   & 1.62 & 1.80     & 2.37 +- 1.02e24   \\
11-10.24 & LT/LOTUS   & 1.63 & 1.79     & 9.44 +- 9.06e23   \\
11-12.21 & LT/LOTUS   & 1.64 & 1.79     & 5.90 +- 5.10e23   \\
11-14.26 & LT/LOTUS   & 1.66 & 1.79     & 9.82 +- 9.14e23   \\
12-03.22 & LT/LOTUS   & 1.80 & 1.74     & 3.42 +- 3.18e23   \\
12-10.28 & LT/LOTUS   & 1.85 & 1.72     & 1.40 +- 1.22e24   \\
		\hline
	\end{tabular}
\end{table}

%\begin{figure}
%	\includegraphics[width=\columnwidth]{trappist-afrho.pdf}
%    \caption{$Af\rho$ measured by TRAPPIST, in $R$-band, within $\rho$ = 10,000 km.}
%    \label{fig:trappist-afrho}
%\end{figure}

\section{Discussion}
\label{sec:discussion}

\begin{figure}
	\includegraphics[width=\columnwidth,trim=40 35 85 80,clip]{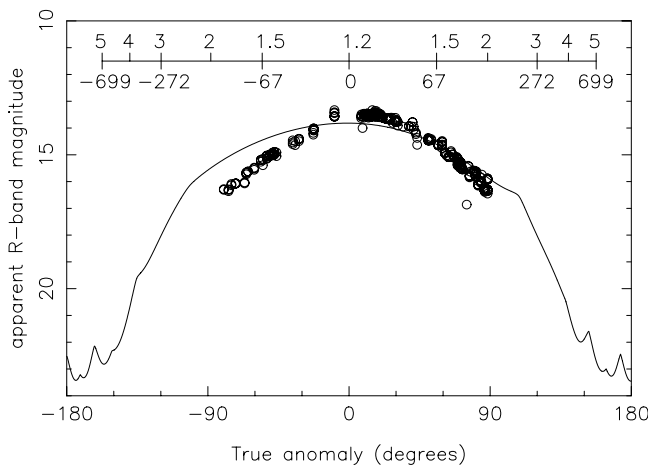}
    \caption{Comparison of photometry (within $\rho = 5$" aperture) with simple thermophysical model. The upper bar shows $r$ and $\Delta T$ (in AU and days, respectively).}
    \label{fig:meech-model}
\end{figure}

If we compare the photometry with predictions from a simple thermophysical model \citep{Meech} we find a reasonable agreement (fig \ref{fig:meech-model}), but there are differences. In this case we plot the photometry measured within a fixed radius aperture ($\rho = 5$") for comparison with the model output, which already includes corrections for changing observation geometry. The match is good close to perihelion, although this model predicts that there should have been more dust lifted by the gas pre-perihelion. It is interesting to compare this with the fits to previous apparitions \citep[fig 9]{Snodgrass2013}, where this model gave a very good fit to all data apart from the few points closest to perihelion. The model required an active surface area of 1.4\% (of the area of a spherical nucleus), and previously needed an enhancement (to 4\%) to match the near perihelion points, while this apparition shows a smoothly varying total brightness through perihelion. As the coverage of the perihelion period is much more dense in our data set than for previous apparitions, this could be used to improve the outgassing model -- it is worth repeating that the model shown in Fig.~\ref{fig:meech-model} is simply an extrapolation of the earlier fit, and has not been adjusted to try to fit the data set plotted here.

%(i.e. comparison with active area model) . Do we see same peak above model around perihelion as previous orbits? If so, when does it turn on/off? compare with illumination of nucleus regions from rosetta.

%- compare CN with Rosina results?? (get CN/H2O ratio?)

Fig \ref{fig:cn_production} indicates that although there is some scatter in the measured CN production rates, Q(CN) was systematically lower in 2015 than during the 1982/83 and 1995/96 post-perihelion apparitions, as measured via narrowband photometry by \citet{Schleicher2006}. It is important to note that the TRAPPIST photometry just after perihelion agrees with the LOTUS spectroscopy. Therefore this appears to be a real change and not caused by any differences between analysing narrow-band aperture photometry and long-slit spectroscopy. However with these data alone we cannot associate this with a secular decrease in outgassing rate. 

It appears that the CN production rate follows a different pattern than the total dust production, both in terms of similarity to previous orbits and the symmetry around the peak -- a longer term view, including observations pre-perihelion and at larger distance post-perihelion with larger aperture telescopes (VLT/FORS) will allow this to be investigated in more detail (Opitom et al, in prep.). The period covered by our TRAPPIST and LOTUS observations corresponds to the time when the total brightness matches the model from gas production (Fig.~\ref{fig:meech-model}), even though they suggest lower total CN production at this time. CN production was even lower pre-perihelion (Opitom et al., in prep), suggesting that the lower-than-predicted brightness there could be related to lower gas production in this period, but CN is a minor component of the coma; total water production measurements from a variety of Rosetta instruments show a more symmetrical pattern that resembles the total dust production seen in Fig.~\ref{fig:timeplot} \citep{Hansen-MNRAS}. Instead, the difference in total CN production may be related to different seasonal illumination of the nucleus.

Although there is some indication that the gas production rate is variable from orbit-to-orbit, the total dust production remains very stable. Further remote observations also support the idea that the activity of the comet is quite stable, in particular as the structure of the coma. Observations from the Wendelstein observatory in Germany show the same large scale structures (jets) as were seen in previous orbits \citep{Boehnhardt-MNRAS,Vincent2013}, indicating that the global activity pattern is similar, and the pole position hasn't significantly changed. Higher resolution polarimetric imaging with the Hubble Space Telescope also reveals the same jet structures \citep{Hadamcik-MNRAS}.

The robotic telescopes provided a very dense monitoring of the comet activity around perihelion, and therefore gave the best chance to detect small outbursts. We do not see any convincing  evidence for outbursts in this data-set, and we did not detect in our $r$-band photometry any of the many small `outbursts' detected by Rosetta \citep{outbursts}. This underlines the fact that ground-based photometry naturally `smears out' the underlying short-timescale activity of the nucleus, due to the photometric apertures containing both the outflowing dust coma generated from the entire nucleus over several hours or days, plus any underlying slow moving gravitationally bound dust particles. Given the previous spacecraft detections of multiple small outbursts i.e. \citet{AHearn2005}, it is possible that the majority of comets undergo continuous small outbursts that fail to be detected even with systematic monitoring as described in this paper.

\section{Conclusions}

We present photometry and spectroscopy of 67P around its 2015 perihelion passage, acquired with robotic telescopes. These telescopes (the 2m LT, 1.2m STELLA, 0.6m TRAPPIST and 1m and 2m telescopes in the LCOGT network) were able to perform very regular observation despite the challenging observing geometry (low solar elongation). We find:

\begin{enumerate}
\item The total brightness of the comet varies smoothly through perihelion, with a peak $\sim$ 2 weeks after closest approach to the Sun.
\item The $R$-band brightness largely follows the prediction from \citet{Snodgrass2013}, indicating that the dust activity level does not change significantly from orbit to orbit.
\item The dust brightness variation is quite symmetrical around its peak, and drops off fairly quickly post perihelion.
\item The gas production (measured in CN) drops off smoothly and slowly post perihelion.
\item There is evidence of a  decrease in the production rate of CN between the 1980's/1990's and the current apparition, {although this needs to be confirmed with observations over a longer period with large telescopes}.
\end{enumerate}

\section*{Acknowledgements}
This work was based, in part, on observations conducted at Canary Island telescopes under CCI ITP programme 2015/06, including use of the Liverpool and STELLA telescopes. 
The Liverpool Telescope is operated on the island of La Palma by Liverpool John
Moores University in the Spanish Observatorio del Roque de los Muchachos of the
Instituto de Astrof\'isica de Canarias with financial support from the UK
Science and Technology Facilities Council (STFC).
The STELLA robotic telescopes in Tenerife are an AIP facility jointly operated by AIP and IAC.
TRAPPIST  is  a  project  funded  by  the  Belgian  Fund  for Scientific
Research (Fonds National de la Recherche Scientifique, F.R.S.-FNRS) under grant
FRFC  2.5.594.09.F,  with  the  participation  of  the  Swiss  National Science
Foundation (SNF). 
CS was supported by an STFC Rutherford fellowship.
CO acknowledges the support of the FNRS. 
EJ is FNRS Research Associate and JM is Research Director of the FNRS.
AF acknowledges support from STFC grant ST/L000709/1.
KM acknowledges support from NSF grant AST1413736.
ACLR acknowledges partial support from CNES.

The full versions of tables 1 and A1 are available in machine readable format online at MNRAS or via the CDS, via anonymous ftp to cdsarc.u-strasbg.fr (130.79.128.5) or via http://cdsarc.u-strasbg.fr/viz-bin/qcat?J/MNRAS/462/S138 

%%%%%%%%%%%%%%%%%%%%%%%%%%%%%%%%%%%%%%%%%%%%%%%%%%



\bibliographystyle{mnras}
\bibliography{67P-perihelion} % if your bibtex file is called example.bib

\begin{thebibliography}{}
\makeatletter
\relax
\def\mn@urlcharsother{\let\do\@makeother \do\$\do\&\do\#\do\^\do\_\do\%\do\~}
\def\mn@doi{\begingroup\mn@urlcharsother \@ifnextchar [ {\mn@doi@}
  {\mn@doi@[]}}
\def\mn@doi@[#1]#2{\def\@tempa{#1}\ifx\@tempa\@empty \href
  {http://dx.doi.org/#2} {doi:#2}\else \href {http://dx.doi.org/#2} {#1}\fi
  \endgroup}
\def\mn@eprint#1#2{\mn@eprint@#1:#2::\@nil}
\def\mn@eprint@arXiv#1{\href {http://arxiv.org/abs/#1} {{\tt arXiv:#1}}}
\def\mn@eprint@dblp#1{\href {http://dblp.uni-trier.de/rec/bibtex/#1.xml}
  {dblp:#1}}
\def\mn@eprint@#1:#2:#3:#4\@nil{\def\@tempa {#1}\def\@tempb {#2}\def\@tempc
  {#3}\ifx \@tempc \@empty \let \@tempc \@tempb \let \@tempb \@tempa \fi \ifx
  \@tempb \@empty \def\@tempb {arXiv}\fi \@ifundefined
  {mn@eprint@\@tempb}{\@tempb:\@tempc}{\expandafter \expandafter \csname
  mn@eprint@\@tempb\endcsname \expandafter{\@tempc}}}

\bibitem[\protect\citeauthoryear{{A'Hearn}, {Schleicher}, {Millis}, {Feldman}
  \& {Thompson}}{{A'Hearn} et~al.}{1984}]{Ahearn84}
{A'Hearn} M.~F.,  {Schleicher} D.~G.,  {Millis} R.~L.,  {Feldman} P.~D.,
  {Thompson} D.~T.,  1984, \mn@doi [\aj] {10.1086/113552}, \href
  {http://ukads.nottingham.ac.uk/cgi-bin/nph-bib_query?bibcode=1984AJ.....89..579A&db_key=AST}
  {89, 579}

\bibitem[\protect\citeauthoryear{{A'Hearn}, {Millis}, {Schleicher}, {Osip}  \&
  {Birch}}{{A'Hearn} et~al.}{1995}]{Ahearn1995}
{A'Hearn} M.~F.,  {Millis} R.~L.,  {Schleicher} D.~G.,  {Osip} D.~J.,   {Birch}
  P.~V.,  1995, \mn@doi [Icarus] {10.1006/icar.1995.1190}, \href
  {http://ukads.nottingham.ac.uk/cgi-bin/nph-bib_query?bibcode=1995Icar..118..223A&db_key=AST}
  {118, 223}

\bibitem[\protect\citeauthoryear{{A'Hearn} et~al.,}{{A'Hearn}
  et~al.}{2005}]{AHearn2005}
{A'Hearn} M.~F.,  et~al., 2005, \mn@doi [Science] {10.1126/science.1118923},
  \href {http://adsabs.harvard.edu/abs/2005Sci...310..258A} {310, 258}

\bibitem[\protect\citeauthoryear{{Barnsley}, {Smith}  \& {Steele}}{{Barnsley}
  et~al.}{2012}]{2012AN....333..101B}
{Barnsley} R.~M.,  {Smith} R.~J.,   {Steele} I.~A.,  2012, \mn@doi
  [Astronomische Nachrichten] {10.1002/asna.201111634}, \href
  {http://adsabs.harvard.edu/abs/2012AN....333..101B} {333, 101}

\bibitem[\protect\citeauthoryear{{Bertin}}{{Bertin}}{2006}]{SCAMP}
{Bertin} E.,  2006, in {Gabriel} C.,  {Arviset} C.,  {Ponz} D.,   {Enrique} S.,
   eds,  Astronomical Society of the Pacific Conference Series Vol. 351,
  Astronomical Data Analysis Software and Systems XV. p.~112

\bibitem[\protect\citeauthoryear{{Bertin} \& {Arnouts}}{{Bertin} \&
  {Arnouts}}{1996}]{SExtractor}
{Bertin} E.,  {Arnouts} S.,  1996, \mn@doi [\aaps] {10.1051/aas:1996164}, \href
  {http://adsabs.harvard.edu/abs/1996A%26AS..117..393B} {117, 393}

\bibitem[\protect\citeauthoryear{{Bilir}, {Ak}, {Karaali}, {Cabrera-Lavers},
  {Chonis}  \& {Gaskell}}{{Bilir} et~al.}{2008}]{2008MNRAS.384.1178B}
{Bilir} S.,  {Ak} S.,  {Karaali} S.,  {Cabrera-Lavers} A.,  {Chonis} T.~S.,
  {Gaskell} C.~M.,  2008, \mn@doi [\mnras] {10.1111/j.1365-2966.2007.12783.x},
  \href {http://adsabs.harvard.edu/abs/2008MNRAS.384.1178B} {384, 1178}

\bibitem[\protect\citeauthoryear{{Boehnhardt}, {Riffeser}, {Kluge}, {Ries},
  {Schmidt}  \& {Hopp}}{{Boehnhardt} et~al.}{2016}]{Boehnhardt-MNRAS}
{Boehnhardt} H.,  {Riffeser} A.,  {Kluge} M.,  {Ries} C.,  {Schmidt} M.,
  {Hopp} U.,  2016, MNRAS, submitted

\bibitem[\protect\citeauthoryear{{Boroson} et~al.,}{{Boroson}
  et~al.}{2014}]{Boroson2014}
{Boroson} T.,  et~al., 2014, in Observatory Operations: Strategies, Processes,
  and Systems V. p. 91491E, \mn@doi{10.1117/12.2054776}

\bibitem[\protect\citeauthoryear{{Brown} et~al.,}{{Brown}
  et~al.}{2013}]{Brown2013}
{Brown} T.~M.,  et~al., 2013, \mn@doi [\pasp] {10.1086/673168}, \href
  {http://adsabs.harvard.edu/abs/2013PASP..125.1031B} {125, 1031}

\bibitem[\protect\citeauthoryear{{Capaccioni} et~al.,}{{Capaccioni}
  et~al.}{2015}]{Capaccioni2015Sci}
{Capaccioni} F.,  et~al., 2015, \mn@doi [Science] {10.1126/science.aaa0628},
  \href {http://adsabs.harvard.edu/abs/2015Sci...347.0628C} {347, 628}

\bibitem[\protect\citeauthoryear{{Farnham}, {Schleicher}  \&
  {A'Hearn}}{{Farnham} et~al.}{2000}]{Farnham2000}
{Farnham} T.~L.,  {Schleicher} D.~G.,   {A'Hearn} M.~F.,  2000, \mn@doi
  [\icarus] {10.1006/icar.2000.6420}, \href
  {http://esoads.eso.org/abs/2000Icar..147..180F} {147, 180}

\bibitem[\protect\citeauthoryear{{Fornasier} et~al.,}{{Fornasier}
  et~al.}{2015}]{Fornasier-AA}
{Fornasier} S.,  et~al., 2015, \aap, \href
  {http://adsabs.harvard.edu/abs/2015arXiv150506888F} {583, A30}

\bibitem[\protect\citeauthoryear{{Giorgini}}{{Giorgini}}{2015}]{horizons}
{Giorgini} J.~D.,  2015, and the JPL Solar System Dynamics Group, NASA JPL
  Horizons On-Line Ephemeris System, <http://ssd.jpl.nasa.gov/?horizons>

\bibitem[\protect\citeauthoryear{{Hadamcik}, {Levasseur-Regourd}, {Hines},
  {Sen}, {Lasue}  \& {Renard}}{{Hadamcik} et~al.}{2016}]{Hadamcik-MNRAS}
{Hadamcik} E.,  {Levasseur-Regourd} A.-C.,  {Hines} D.~C.,  {Sen} A.,  {Lasue}
  J.,   {Renard} J.-B.,  2016, MNRAS, submitted

\bibitem[\protect\citeauthoryear{{Hansen} et~al.,}{{Hansen}
  et~al.}{2016}]{Hansen-MNRAS}
{Hansen} K.~C.,  et~al., 2016, MNRAS, submitted

\bibitem[\protect\citeauthoryear{{Haser}}{{Haser}}{1957}]{Haser1957}
{Haser} L.,  1957, Bulletin de la Societe Royale des Sciences de Liege, \href
  {http://esoads.eso.org/abs/1957BSRSL..43..740H} {43, 740}

\bibitem[\protect\citeauthoryear{{Holmberg}, {Flynn}  \&
  {Portinari}}{{Holmberg} et~al.}{2006}]{Holmberg2006}
{Holmberg} J.,  {Flynn} C.,   {Portinari} L.,  2006, \mn@doi [\mnras]
  {10.1111/j.1365-2966.2005.09832.x}, \href
  {http://adsabs.harvard.edu/abs/2006MNRAS.367..449H} {367, 449}

\bibitem[\protect\citeauthoryear{{Jehin} et~al.,}{{Jehin}
  et~al.}{2011}]{TRAPPIST}
{Jehin} E.,  et~al., 2011, The Messenger, \href
  {http://esoads.eso.org/abs/2011Msngr.145....2J} {145, 2}

\bibitem[\protect\citeauthoryear{{Jenness} \& {Economou}}{{Jenness} \&
  {Economou}}{2015}]{Jenness2015}
{Jenness} T.,  {Economou} F.,  2015, \mn@doi [Astronomy and Computing]
  {10.1016/j.ascom.2014.10.005}, \href
  {http://adsabs.harvard.edu/abs/2015A%26C.....9...40J} {9, 40}

\bibitem[\protect\citeauthoryear{{Jester} et~al.,}{{Jester}
  et~al.}{2005}]{2005AJ....130..873J}
{Jester} S.,  et~al., 2005, \mn@doi [\aj] {10.1086/432466}, \href
  {http://adsabs.harvard.edu/abs/2005AJ....130..873J} {130, 873}

\bibitem[\protect\citeauthoryear{{Le Roy} et~al.,}{{Le Roy}
  et~al.}{2015}]{LeRoy2015}
{Le Roy} L.,  et~al., 2015, \mn@doi [A\&A] {10.1051/0004-6361/201526450}, 583,
  A1

\bibitem[\protect\citeauthoryear{{Luspay-Kuti} et~al.,}{{Luspay-Kuti}
  et~al.}{2015}]{LuspayKuti2015}
{Luspay-Kuti} A.,  et~al., 2015, \mn@doi [A\&A] {10.1051/0004-6361/201526205},
  583, A4

\bibitem[\protect\citeauthoryear{{Marcus}}{{Marcus}}{2007}]{Marcus2007}
{Marcus} J.~N.,  2007, International Comet Quarterly, \href
  {http://esoads.eso.org/abs/2007ICQ....29..119M} {29, 119}

\bibitem[\protect\citeauthoryear{{Meech} \& {Svore{\v n}}}{{Meech} \& {Svore{\v
  n}}}{2004}]{Meech}
{Meech} K.~J.,  {Svore{\v n}} J.,  2004, in {Festou} M.~C.,  {Keller} H.~U.,
  {Weaver} H.~A.,  eds, , Comets II.
University of Arizona Press, pp 317--335

\bibitem[\protect\citeauthoryear{{Opitom}, {Jehin}, {Manfroid},
  {Hutsem{\'e}kers}, {Gillon}  \& {Magain}}{{Opitom} et~al.}{2015}]{Opitom2015}
{Opitom} C.,  {Jehin} E.,  {Manfroid} J.,  {Hutsem{\'e}kers} D.,  {Gillon} M.,
   {Magain} P.,  2015, \mn@doi [\aap] {10.1051/0004-6361/201424582}, \href
  {http://esoads.eso.org/abs/2015A%26A...574A..38O} {574, A38}

\bibitem[\protect\citeauthoryear{{Piascik}, {Steele}, {Bates}, {Mottram},
  {Smith}, {Barnsley}  \& {Bolton}}{{Piascik} et~al.}{2014}]{Piascik2014}
{Piascik} A.~S.,  {Steele} I.~A.,  {Bates} S.~D.,  {Mottram} C.~J.,  {Smith}
  R.~J.,  {Barnsley} R.~M.,   {Bolton} B.,  2014, in Ground-based and Airborne
  Instrumentation for Astronomy V. p. 91478H, \mn@doi{10.1117/12.2055117}

\bibitem[\protect\citeauthoryear{{Roeser}, {Demleitner}  \&
  {Schilbach}}{{Roeser} et~al.}{2010}]{2010AJ....139.2440R}
{Roeser} S.,  {Demleitner} M.,   {Schilbach} E.,  2010, \mn@doi [\aj]
  {10.1088/0004-6256/139/6/2440}, \href
  {http://adsabs.harvard.edu/abs/2010AJ....139.2440R} {139, 2440}

\bibitem[\protect\citeauthoryear{{Schleicher}}{{Schleicher}}{2006}]{Schleicher2006}
{Schleicher} D.~G.,  2006, \mn@doi [\icarus] {10.1016/j.icarus.2005.11.014},
  \href {http://esoads.eso.org/abs/2006Icar..181..442S} {181, 442}

\bibitem[\protect\citeauthoryear{{Schleicher}}{{Schleicher}}{2010}]{2010AJ....140..973S}
{Schleicher} D.~G.,  2010, \mn@doi [\aj] {10.1088/0004-6256/140/4/973}, \href
  {http://adsabs.harvard.edu/abs/2010AJ....140..973S} {140, 973}

\bibitem[\protect\citeauthoryear{{Schleicher}, {Millis}  \&
  {Birch}}{{Schleicher} et~al.}{1998}]{Schleicher1998}
{Schleicher} D.~G.,  {Millis} R.~L.,   {Birch} P.~V.,  1998, \mn@doi [\icarus]
  {10.1006/icar.1997.5902}, \href
  {http://esoads.eso.org/abs/1998Icar..132..397S} {132, 397}

\bibitem[\protect\citeauthoryear{{Snodgrass}, {Tubiana}, {Bramich}, {Meech},
  {Boehnhardt}  \& {Barrera}}{{Snodgrass} et~al.}{2013}]{Snodgrass2013}
{Snodgrass} C.,  {Tubiana} C.,  {Bramich} D.~M.,  {Meech} K.,  {Boehnhardt} H.,
    {Barrera} L.,  2013, \mn@doi [\aap] {10.1051/0004-6361/201322020}, \href
  {http://esoads.eso.org/abs/2013A%26A...557A..33S} {557, A33}

\bibitem[\protect\citeauthoryear{{Snodgrass} et~al.,}{{Snodgrass}
  et~al.}{2016}]{Snodgrass2016}
{Snodgrass} C.,  et~al., 2016, \mn@doi [\aap] {10.1051/0004-6361/201527834},
  \href {http://adsabs.harvard.edu/abs/2016A%26A...588A..80S} {588, A80}

\bibitem[\protect\citeauthoryear{{Steele} et~al.,}{{Steele}
  et~al.}{2004}]{Steele2004}
{Steele} I.~A.,  et~al., 2004, in {Oschmann} Jr. J.~M.,  ed.,  \procspie Vol.
  5489, Ground-based Telescopes. pp 679--692, \mn@doi{10.1117/12.551456}

\bibitem[\protect\citeauthoryear{{Steele}, {Mottram}, {Smith}  \&
  {Barnsley}}{{Steele} et~al.}{2014}]{steele2014}
{Steele} I.~A.,  {Mottram} C.~J.,  {Smith} R.~J.,   {Barnsley} R.~M.,  2014, in
  High Energy, Optical, and Infrared Detectors for Astronomy VI. p. 915428,
  \mn@doi{10.1117/12.2056602}

\bibitem[\protect\citeauthoryear{{Steele} et~al.,}{{Steele}
  et~al.}{2016}]{lotus}
{Steele} I.~A.,  et~al., 2016, \mn@doi [\mnras] {10.1093/mnras/stw1287}, \href
  {http://adsabs.harvard.edu/abs/2016MNRAS.tmp..956S} {460, 4268}

\bibitem[\protect\citeauthoryear{{Strassmeier} et~al.,}{{Strassmeier}
  et~al.}{2004}]{STELLA2}
{Strassmeier} K.~G.,  et~al., 2004, \mn@doi [Astronomische Nachrichten]
  {10.1002/asna.200410273}, \href
  {http://cdsads.u-strasbg.fr/abs/2004AN....325..527S} {325, 527}

\bibitem[\protect\citeauthoryear{{Strassmeier} et~al.,}{{Strassmeier}
  et~al.}{2010}]{STELLA1}
{Strassmeier} K.~G.,  et~al., 2010, \mn@doi [Advances in Astronomy]
  {10.1155/2010/970306}, \href
  {http://cdsads.u-strasbg.fr/abs/2010AdAst2010E..19S} {2010, 970306}

\bibitem[\protect\citeauthoryear{{Vincent}, {Lara}, {Tozzi}, {Lin}  \&
  {Sierks}}{{Vincent} et~al.}{2013}]{Vincent2013}
{Vincent} J.-B.,  {Lara} L.~M.,  {Tozzi} G.~P.,  {Lin} Z.-Y.,   {Sierks} H.,
  2013, \mn@doi [\aap] {10.1051/0004-6361/201219350}, \href
  {http://esoads.eso.org/abs/2013A%26A...549A.121V} {549, A121}

\bibitem[\protect\citeauthoryear{{Vincent}, {A'Hearn}, {Lin}, {El-Maary},
  {Pajola}  \& {Osiris Team}}{{Vincent} et~al.}{2016}]{outbursts}
{Vincent} J.-B.,  {A'Hearn} M.~F.,  {Lin} Z.-Y.,  {El-Maary} M.~R.,  {Pajola}
  M.,   {Osiris Team} 2016, MNRAS, submitted

\bibitem[\protect\citeauthoryear{{Zacharias}, {Finch}, {Girard}, {Henden},
  {Bartlett}, {Monet}  \& {Zacharias}}{{Zacharias}
  et~al.}{2013}]{Zacharias2013UCAC4}
{Zacharias} N.,  {Finch} C.~T.,  {Girard} T.~M.,  {Henden} A.,  {Bartlett}
  J.~L.,  {Monet} D.~G.,   {Zacharias} M.~I.,  2013, \mn@doi [\aj]
  {10.1088/0004-6256/145/2/44}, \href
  {http://adsabs.harvard.edu/abs/2013AJ....145...44Z} {145, 44}

\makeatother
\end{thebibliography}

%%%%%%%%%%%%%%%%%%%%%%%%%%%%%%%%%%%%%%%%%%%%%%%%%%

%%%%%%%%%%%%%%%%% APPENDICES %%%%%%%%%%%%%%%%%%%%%

\appendix

\section{Photometry}

This appendix gives all $R$-band photometry from the robotic telescopes mentioned in this paper. All values are based on an aperture with radius $\rho$ = 10,000 km at the distance of the comet. Where the telescopes used an SDSS-$r$ filter, a colour correction of -0.2105 has been applied. Heliocentric distance ($r$) and time from perihelion ($\Delta T$) are given, together with the apparent magnitude as well as the magnitude corrected to unit geocentric distance and zero phase angle, using $\beta = 0.02$ mag.~deg.$^{-1}$. The $Af\rho$ quantity is given in cm, \emph{without} any correction for phase angle in this case.

\begin{table}
% info from TRAPPIST and LOTUS, sensible columns TBC.
% uncertainties need to be adjusted
	\centering
	\caption{$R$-band Photometry. Measured within an aperture with radius $\rho$ 10,000 km. {Full table available online, first 5 rows given as an example.}}
	\label{tab:phot}
	\begin{tabular}{lcccccl}
		\hline
UT Date   & $r$  & $\Delta T$ & mag & R(r,1,0) & $Af\rho$ &  Tel./inst.   \\
             &             (AU) & (days)     & & & (cm)     \\
		\hline
04-18.41 &  1.83 & -116.68 & 16.280 & 13.859 &  94.3 & TRAPPIST \\
04-18.42 &  1.83 & -116.67 & 16.310 & 13.889 &  91.7 & TRAPPIST \\
04-25.41 &  1.78 & -109.68 & 16.230 & 13.846 &  87.4 & TRAPPIST \\
04-25.42 &  1.78 & -109.67 & 16.160 & 13.776 &  93.2 & TRAPPIST \\
04-25.42 &  1.78 & -109.67 & 16.210 & 13.826 &  89.0 & TRAPPIST \\
... & ... & ... & ... & ... & ... & ... \\
\hline
\end{tabular}
\end{table}

%%%%%%%%%%%%%%%%%%%%%%%%%%%%%%%%%%%%%%%%%%%%%%%%%%

% Don't change these lines
\bsp	% typesetting comment
\label{lastpage}
\end{document}